\documentclass[12pt, letterpaper]{article}

\usepackage{amsmath}
\usepackage{amsfonts}
\usepackage{amssymb}
\usepackage{graphicx}

\usepackage{color}

\usepackage{amsmath, amssymb, graphics, epsfig, graphicx}
\usepackage{epsf}
\usepackage{epstopdf}
\usepackage {amssymb}
\newcommand{\nc}{\newcommand}
\nc{\ba}{\begin{eqnarray}}
\nc{\ea}{\end{eqnarray}}
\newcommand\be{\begin{equation}}
\newcommand\ee{\end{equation}}

\newcommand{\calR}{{\cal{R}}}

\newcommand{\bfx}{{\bf{x}}}

\begin{document}

\vspace{5mm}
\vspace{0.5cm}
\begin{center}

\def\thefootnote{\fnsymbol{footnote}}

{\Large  {\bf  Instabilities in Mimetic Matter Perturbations} }
\\[0.5cm]

{  Hassan Firouzjahi$^{1}\footnote{firouz@ipm.ir }$,  Mohammad Ali Gorji$^{1}\footnote{gorji@ipm.ir}$, Seyed Ali Hosseini Mansoori$^{ 2, 1}\footnote{ shosseini@shahroodut.ac.ir, shossein@ipm.ir  }$}
\\[0.5cm]

{\small \textit{$^1$School of Astronomy, Institute for Research in Fundamental Sciences (IPM) \\ P.~O.~Box 19395-5531, Tehran, Iran
}}\\

{\small \textit{$^2$
Physics Department, Shahrood University of Technology,\\ P.O.Box 3619995161 Shahrood, Iran }}

\end{center}

\vspace{.8cm}

\hrule \vspace{0.3cm}


\begin{abstract}
We study cosmological perturbations in mimetic matter scenario with a general higher derivative function. 
We calculate the quadratic action and show that  both the kinetic term and the gradient term have the wrong sings. We perform the analysis in both comoving and Newtonian  gauges and confirm that the Hamiltonians and the associated instabilities are consistent with each  other in both gauges.  The existence of instabilities is independent of the specific 
form of  higher derivative function which generates  gradients for mimetic field perturbations.  It is verified that the ghost instability in mimetic perturbations is not associated with the higher derivative instabilities such as the Ostrogradsky ghost. 
\end{abstract}
\vspace{0.5cm} \hrule
\def\thefootnote{\arabic{footnote}}
\setcounter{footnote}{0}
\newpage
\section{Introduction}

The $\Lambda CDM$ model of cosmology based on cosmological constant  and cold dark matter as the main sources of energy in evolution  of cosmos was a successful paradigm  in explaining a host of cosmological observations. Despite its observational success, the $\Lambda CDM$ model  is still a phenomenological model. There are promising hopes to embed  cold dark matter in theories beyond the Standard Model (SM) of particle physics. However, it is still an open question whether modifying gravity can play the role of dark matter.

Recently the ``Mimetic Matter'' scenario has been proposed as an alternative to dark matter \cite{Chamseddine:2013kea, Chamseddine:2014vna}. In this picture, the ``physical'' metric $g_{\mu \nu}$ is related to a scalar field and an auxiliary metric $\tilde g_{\mu \nu} $ via  $g_{\mu \nu} = - (\tilde g^{\alpha \beta } \partial_\alpha \phi \partial_\beta \phi ) \tilde g_{\mu \nu} $ (we use the metric sign convention $(-, +, +, +) )$. 
The total action is written in terms of the usual Einstein-Hilbert term plus the contribution of any matter, say from the SM sector, which couples to $g_{\mu \nu}$. The resulting gravitational fields equations describe the usual Einstein's equations plus the contribution of the mimetic field which appears as a new source of  stress energy tensor. More specifically, the degree of freedom associated with the mimetic field is encoded into the longitudinal mode of gravity which becomes dynamical even in the absence of conventional SM fields. It is shown in \cite{Chamseddine:2013kea} that this additional  mode of gravitation can mimic the roles of dark matter. From the identification of metric $g_{\mu \nu}$ in terms of $ \tilde g_{\mu \nu}$
one obtains the mimetic constraint 
\ba
\label{mimetic-const.}
g^{\alpha \beta } \partial_\alpha \phi \partial_\beta \phi  =-1 
\ea
which should be implemented in the dynamics of the mimetic field. 

One can view the mimetic scenario as a  particular limit of the general disformal transformation involving the auxiliary metric $\tilde g_{\mu \nu} $, $\partial_\mu \phi \partial_\nu \phi$ and the scalar quantity  $\tilde X\equiv \tilde g^{\mu \nu} \partial_\mu \phi \partial_\nu \phi$. It is shown that the mimetic scenario is a singular limit of the general disformal transformation in  which the  operation  expressing $ g_{\mu \nu}$ in terms of $ \tilde g_{\mu \nu}$ and $\phi$ is not invertible \cite{Deruelle:2014zza, Arroja:2015wpa, Domenech:2015tca}.

The cosmological perturbations in the original mimetic scenario was performed in \cite{Chamseddine:2014vna} in which it is found that the $\delta \phi$ fluctuations in Newtonian gauge  has zero propagation velocity, i.e. it has zero sound speed $c_s=0$.  This may rise to caustics in perturbations associated with dark matter. In addition, the usual notion of quantum wave associated with  $\delta \phi$ perturbations does not exist as there is no propagating degrees of freedom  for $\delta \phi(k)$ modes in Fourier space. To remedy these issues, it is suggested in \cite{Chamseddine:2014vna} to supplement the simple mimetic scenario with higher derivative terms 
such as $\gamma (\Box \phi)^2$ to generate gradient terms for $\delta \phi$ excitations yielding a proper propagating degrees of freedom with a non-zero sound speed. More recently, mimetic scenario with
a general  higher derivative function in the form $f( \Box \phi)$ has been studied in \cite{Chamseddine:2016uef} and \cite{Chamseddine:2016ktu} which is argued to play important roles is resolving the cosmological and black hole singularities. 
  
In this paper we revisit the question of cosmological perturbations in the mimetic setup with the  general higher derivative function  $f( \Box \phi)$ as included in \cite{Chamseddine:2016uef} and \cite{Chamseddine:2016ktu} (see also \cite{Arroja:2015yvd, Cognola:2016gjy} for cosmological perturbations of mimetic Horndeski setup). 
Our goal is to see if the theory is stable under small perturbations.  As we shall see, the kinetic term and the gradient term in the action have the wrong signs and the scenario  may suffer from instabilities.   

The paper is organized as follows.  In section \ref{setup-sec} we present the mimetic setup in some details.   In section \ref{comoving-sec} we  study  perturbations in comoving gauge where it is easy to see the existence of  instabilities. To confirm the existence of instabilities, in section \ref{Newton-sec} we study perturbations in Newtonian gauge  which is also the gauge used in  \cite{Chamseddine:2014vna}.  The  Hamiltonian analysis of the perturbations are presented 
in section \ref{Hamiltonian-sec}  followed by Discussions in section \ref{summary-sec}.  Some technical formulas are relegated to the Appendix.

\section{The Mimetic setup}
\label{setup-sec}

In this section we review the mimetic matter scenario in some details. The action containing the metric $g_{\mu \nu}$ and the mimetic field is given by
\ba
\label{action0}
S= \int d^4 x  \sqrt{-g} \left[ \frac{M_P^2}{2} R + \lambda \left( g^{\alpha \beta } \partial_\alpha \phi \partial_\beta \phi + 1 \right) + f (\chi) - V(\phi)
\right]
\ea
in which $M_P$ is the reduced Planck mass.  The term $\lambda$ in the above action is a Lagrange multiplier which is added to enforce the mimetic constraint  Eq. (\ref{mimetic-const.}) \cite{Golovnev:2013jxa, Barvinsky:2013mea, Hammer:2015pcx}, see also \cite{Lim:2010yk} for a somewhat similar action with a Lagrange multiplier but in a different setup. 
Also  we have added the potential term $V(\phi)$ for generality, though its specific form  does not affect the stability analysis. The presence of potential may be important if one wants to obtain inflation \cite{Chamseddine:2014vna} or bounce \cite{Ijjas:2016pad} from the mimetic scenario. Finally, the term $f (\chi)$  with 
 \ba
 \chi \equiv \Box \phi
 \ea
 represents the higher derivative term added to the original mimetic model to generate a non-zero sound speed for mimetic field perturbations. The particular case of $f(\chi)
 =\gamma \chi^2/2$ was considered in the analysis of \cite{Ramazanov:2016xhp}  and \cite{Ijjas:2016pad} where the former reference studied the link between the mimetic scenario and  the IR limit of the projectable Horava-Lifshitz gravity while the latter reference  studied the question of ghost and gradient instabilities in bouncing cosmology within the setup of mimetic matter. Our work generalize their  analysis 
for arbitrary form of $f(\chi)$ with emphasis on expanding FRW background.

Varying the action with respect to $g_{\mu \nu}$, one obtains the Einstein fields equations 
$M_P^2 G_{\mu \nu} = T_{\mu \nu} $ in which $G_{\mu \nu}$ is the Einstein tensor and $T_{\mu \nu}$ is the effective energy momentum tensor associated with the mimetic field
\begin{equation}
\label{Tmu-nu}
T^{\mu }_{\nu }=-2\lambda {{\partial }^{\mu }}\phi {{\partial }_{v}}
\phi +\left( {{\partial }^{\mu }}f_{\chi}{{\partial }_{\nu }}\phi +
{{\partial }^{\mu }}\phi {{\partial }_{\nu }}f_{\chi} \right)+
\delta _{\nu }^{\mu }\left( f-\chi f_{\chi }-V-{{g}^{\alpha \beta }}{{
\partial }_{\alpha }}f_{\chi}{{\partial }_{\beta }}\phi  \right) \, ,
\end{equation}
in which $f_{\chi} \equiv \partial f/\partial \chi$.  Note that there is another contribution in the form 
$\delta^\mu_\nu ( g^{\alpha \beta } \partial_\alpha \phi \partial_\beta \phi + 1 )$ in $T^{\mu}_{\nu}$ which vanishes because of the constraint Eq. (\ref{mimetic-const.}) and was not included in Eq. (\ref{Tmu-nu}). Note that the energy momentum tensor given above can not be cast into the form of the energy momentum tensor of a perfect fluid. This is demonstrated in  \cite{Mirzagholi:2014ifa} in which it is shown that there is an energy flow and vorticity for the mimetic matter with $f_{\chi \chi} \neq0$. The fact that the energy momentum tensor of the mimetic setup does not have the form of a perfect fluid may be helpful in curing some phenomenological problems associated with the cold dark matter model \cite{Capela:2014xta, Ramazanov:2015pha}. For other works on various aspects of mimetic scenario see 
\cite{Nojiri:2014zqa,Leon:2014yua,Sebastiani:2016ras, Momeni:2014qta,Haghani:2015iva}.

In addition, varying the action with respect to $\phi$ yields the modified Klein-Gordon equation
\begin{equation}
\label{KG-eq}
\frac{1}{\sqrt{-g}}{{\partial }_{\mu }}\Big( \sqrt{-g}\left( 2\lambda {{\partial }^{\mu }}\phi -{{\partial }^{\mu }}f_{\chi}\right) \Big)+\frac{\partial V\left( \phi  \right)}{\partial \phi }=0 \, .
\end{equation}

In an FRW background with the metric  
\ba
ds^2 = - dt^2 + a(t)^2 d \bfx^2 \,,
\ea
the background Einstein equations are
\ba
3 M_P^2 H^2 = V- 2 \lambda - ( f + 3 H f_{\chi } + 3 \dot H f_{\chi \chi})  \, ,
\ea
and
\ba
M_P^2 (2 \dot H + 3 H^2 ) = V- ( f + 3 H f_{\chi } - 3 \dot H f_{\chi \chi}) \, ,
\ea
in which $H = \dot a(t)/a(t)$ is the Hubble expansion rate.

Combining the above equations, one can solve for  $\lambda$ as follows
\ba
\label{lambda-eq0}
\lambda = ( M_P^2 - 3 f_{\chi \chi}) \dot H = \epsilon  ( 3 f_{\chi \chi} - M_P^2) H^2 \, ,
\ea
in which $\epsilon $ is defined (like the slow-roll parameter in inflation) as $\epsilon \equiv - \dot H/H^2$.
In our analysis below, we are mainly interested in an expanding FRW background in which the null energy condition is valid with $\dot H <0$ so $\epsilon >0$. Alternatively, in general and in the presence of perturbations, one can obtain $\lambda$ from the trace of $T^\mu_\nu$ combined with Einstein's equation as follows
\ba
\label{lambda-eq}
\lambda= \frac{1}{2} \Big( G + 4 (V+ \chi f_{\chi } - f) + 2 \partial^\mu \phi \partial_\mu f_{\chi }
\Big) \, ,
\ea
in which $G$ is the trace of the Einstein tensor.  

At the background level, the constraint Eq. (\ref{mimetic-const.}) implies $\dot \phi^2 =1$, which  without loss of generality we can take $\dot \phi=1$ and $\phi = t$. In addition, at the background level we have  $\chi = \Box \phi = -3 H$.

For the simple mimetic setup with $V(\phi) = f(\chi)=0$, from the energy momentum tensor   Eq. (\ref{Tmu-nu}), the energy density $\rho$ and the pressure $P$ are obtained to be  $\rho = -T^0_0 = - 2 \lambda = - \dot H$ and $P=0$. On the other hand, from Eq. (\ref{KG-eq}) one obtains $\partial_t ( \rho a^3)=0$ which is nothing but the evolution of mimetic matter, mimicking the role of cold dark matter with 
zero pressure. 

In our analysis below, we assume $f(\chi)\neq 0$ so the mimetic setup  as suggested in \cite{Chamseddine:2014vna, Chamseddine:2016uef, Chamseddine:2016ktu} is endowed with higher derivative terms. As we commented before, this is necessary to generate a non-zero sound speed for the $\delta \phi$ perturbations and also  to make the quantum origins of  $\delta \phi$ perturbations well-defined.  Having said this, we also comment about the perturbation analysis for the simple mimetic setup 
with $V(\phi) = f(\chi)=0$. 

Before proceeding to perturbation analysis, let us evaluate $T^0_i$ which is used to read off the fluid's velocity potential $\delta u$. From Eq. (\ref{Tmu-nu}) and noting that $\dot \phi =1$, to first order in perturbations we obtain 
\ba
\label{T0-i}
\delta T^0_i = (2 \lambda - f_{\chi \chi}\dot \chi ) \partial_i \delta \phi - f_{\chi \chi} \partial_i \delta \chi \, .
\ea 
Now using the usual convention that $\delta T^0_i$ is related to comoving velocity via
$\delta T^0_i = ( \rho + P) \partial_ i \delta u$, the  velocity
potential to first order in perturbations is obtained to be\footnote{As discussed in \cite{Weinberg:2008zzc}
the identification $\delta T^0_i = ( \rho + P) \partial_ i \delta u$ can be defined for an imperfect fluid too. }
\ba
(\rho + P) \delta u = ( 2 M_P^2 - 3 f_{\chi \chi}) \dot H \delta \phi - f_{\chi \chi} \delta \chi \, .
\ea
In conventional scalar field theories with no higher derivative terms, the comoving hypersurface defined via $\delta u=0$ is identical to hypersurface $\delta \phi=0$. However, in the presence of higher derivative terms $\delta \chi$ also contributes to velocity potential and, technically speaking, 
the hypersurface $\delta \phi=0$ is not the usual comoving hypersurface.

\section{Comoving gauge}
\label{comoving-sec}

In this section we present the analysis of cosmological perturbations in comoving gauge.  As mentioned before, this gauge is particularly convenient to see the existence of the ghost and the gradient instabilities. To simplify the analysis, from now on we set  $M_{P}=1$. 

Going to ADM formalism, the decomposition of metric components  are given by
\begin{equation}\label{ADM-metric}
ds^2=-N^2dt^2+h_{ij}(dx^i+N^i dt)(dx^j+N^j dt) ~.
\end{equation}
In standard general relativity theory the advantage with this decomposition is that the functions $N$ and $N^i$ appears as constraints with no time derivatives so they can be substituted in the action upon  solving the  constraints equations which are algebraic.  However, as we shall see, in mimetic matter scenario with higher derivative term $f(\chi)$, the lapse function $N$ becomes dynamical.  

The spatial metric $h_{ij}$ has two scalar perturbations.  After killing one scalar degree of freedom in $h_{ij}$ (the one involving $\partial_{i} \partial_{j} $ operating on a scalar), 
we set $h_{ij}$ to have the diagonal form  
\begin{equation}\label{hij}
h_{ij}=a^2 e^{2 \psi}\delta_{ij} ~,
\end{equation}
in which $\psi$ is a scalar perturbation measuring the three-dimensional spatial curvature perturbations.  

As usual, the quantity 
\ba
\calR \equiv \psi -\frac{H}{\dot \phi} \delta \phi  \, ,
\ea
is gauge invariant. This quantity in \cite{Ijjas:2016pad} is interpreted as comoving curvature perturbation. 
However, in conventional perturbation theory, the comoving curvature perturbation is defined via
\ba
\calR_c \equiv \psi + H \delta u 
\ea
in which $\delta u$ is the velocity potential as given by $\delta T^0_i = ( \rho + P) \partial_ i \delta u$.  
As we commented at the end of last section, when $f_{\chi \chi} \neq 0$, the hypersurface $\delta \phi=0$ does not coincide with the hypersurface $\delta u=0$ so $\calR \neq \calR_c$. 
Having said this, we follow \cite{Ijjas:2016pad} and define 
the hypersurface $\delta \phi=0$ as the comoving surface and set $\psi = \calR$ in the following analysis. Note that the choice of comoving gauge is always possible, since the vector $\partial_{\mu} \phi$ is timelike as enforced by the mimetic constraint Eq. (\ref{mimetic-const.}) so one can always go to the frame in which $\delta \phi=0$. 

In ADM decomposition, the action takes the following form
\begin{equation} \label{action}
S=\frac{1}{2}\int dt d^3x N\sqrt{h} \Big(^{(3)}R
+N^{-2}(E_{ij}E^{ij}-E^2)\Big)
+\int dt d^3x N\sqrt{h} \Big(\lambda(X+1)+f-V\Big)\,,
\end{equation}
in which we have defined $X\equiv g^{\mu\nu}
\partial_\mu\phi\partial_\nu \phi$, $^{(3)}R$ is the three-dimensional spatial curvature associated with the 
metric $h_{ij}$ and $E_{ij}$ is related to  the extrinsic curvature $K_{ij}$ via $E_{ij} = N K_{ij}$  so
\begin{equation}
E_{ij}=\frac{1}{2}(\dot{h}_{ij}-\nabla_i N_j-\nabla_j N_i) ~.
\label{Edef}
\end{equation}
In addition  $E= E^i_i$ in which the spatial indices are raised and lowered with the metric $h_{ij}$. 

Similar to analysis in \cite{Maldacena:2002vr, Chen:2006nt, 
Seery:2005wm}, the equations of motion for $N$ and 
$N^i$ respectively are
\ba
\label{eomN}
^{(3)}R-N^{-2}(E_{ij}E^{ij}-E^2)
+2\lambda(X+1)- 4X \lambda-2V
\\ \nonumber
~~~~~~~~~~~~~~ + 2f-4\chi {f_{\chi}}-2N^{-3}f_{\chi}\left( \dot{N}+{N^i}{
\partial_i}N\right) = 0 \,,
\ea
and 
\ba
\label{Nieom}
\nabla_j(N^{-1}E_i^j)-\nabla_i(N^{-1}E) 
&=& N^{-1} f_{\chi \chi} {\partial}_{i}\chi 
- 2N^{-2}f_{\chi }{\partial}_i N ~.
\ea
Note that in  the ADM decomposition, the form of $\chi$ is given by 
\begin{equation}
\chi =\square \phi =\frac{1}{{{N}^{2}}}\Big( \frac{{\dot{N}}}{N}+\frac{{{N}^{i}}{{\partial }_{i}}N}{N}+{{\partial }_{i}}{{N}^{i}}+\frac{{{N}^{i}}}{2}\frac{{{\partial }_{i}}h}{h}-\frac{{\dot{h}}}{2h} \Big) \, ,
\end{equation}
which will be useful to calculate $\delta\chi$ perturbatively. 

Since we are interested in scalar perturbation, we consider first order scalar perturbation in metric such that
\ba\label{ADM-scalar-pert}
N=1+\alpha, \quad  N^i=\partial^i \beta,  \quad 
\lambda=\bar \lambda+\lambda^{(1)}\,, 
\ea
where $\bar\lambda$ denotes the background value of  the auxiliary field $\lambda$ and $\lambda^{(1)}$ is its first order perturbation.

Substituting the above power expansion into the 
equations of motion (\ref{eomN}) and  (\ref{Nieom})
for $N$ and $N^i$ and imposing the mimetic 
constraint $X=-1$, at first order in $\calR$, we obtain 
\begin{eqnarray} 
\label{solution1}
\alpha = 0 \,,  \quad 
\beta=\big(3-  \frac{2}{f_{\chi\chi}} \big)
{{\partial }^{-2}}\dot{\calR} ~, \quad 
{\lambda }^{(1)}=\Big(\frac{2(2f_{\chi\chi}-1)}{
a^2(-2+3f_{\chi\chi})}\Big)\partial^2\calR+2H
\big(3- \frac{1}{f_{\chi\chi}}\big)\,\dot{\calR }\,.
\end{eqnarray}
In deriving these results, it is assumed that $f_{\chi \chi} \neq0$. The simple case $f=0$ is discussed separately at the end of this section. 

Plugging the above values into the action and after some integration by parts the quadratic action in comoving gauge  for $\calR$ is obtained to be  
\ba
\label{action-R}
S^{(2)}_{\rm com}=\int{dt\, {{d}^{3}}x\, {{a}^{3}} \Big[ \big(3- \frac{2}{f_{\chi \chi}}\big){{{\dot{\calR }}}^{2}}+\frac{{{\left( \partial \calR \right)}^{2}}}{{{a}^{2}}} \Big]}\,.
\ea
In the special case of $f(\chi)= \gamma \chi^2/2$ the above action coincides with the result of \cite{Ijjas:2016pad}. 

Varying the action (\ref{action-R}) with respect to $\calR$ yields the perturbed equation of motion for 
$\calR$
\ba
\label{mimetic-eq-R}
\partial_{t } \Big( a^{3}   \big( 3- \frac{2}{f_{\chi \chi}}\big)\dot{\calR } \Big) + \frac{\partial^{2} \calR}{a^{2}} =0 \, .
\ea
From this equation of motion the sound speed of scalar perturbations is obtained to be 
\ba
\label{cs-def} 
c_s^2 = \frac{f_{\chi \chi}}{2- 3 f_{\chi \chi}} \, .
\ea
This value of $c_{s}^{2}$ is also verified from the equation of motion in Newtonian gauge.

To study the stability of the system as usual one has to construct the Hamiltonian of the system. Going to Fourier\footnote{For the sake of simplicity, when going  to Fourier space we  do not write the dependence of perturbations on Fourier wave number $k$  so $\calR (k)$ is simply denoted by $\calR$ and so on.} space and working with conformal time $d\tau=dt/a(t)$ the conjugate momentum associated with 
$\calR$ from the action (\ref{action-R}) is given by $\Pi_{\calR}=\partial {\cal L}_{\rm com}/\partial{\calR'}
=-2c_s^{-2}a^2{\calR'}$ in which ${\cal L}_{\rm com}$ is the Lagrangian density from action (\ref{action-R}) and  a prime denotes the derivative with respect to conformal time. Correspondingly, the Hamiltonian (actually the Hamiltonian density in Fourier space) constructed from the Lagrangian of action (\ref{action-R})  is given by 
\ba
\label{Hamiltonian-com-conformal}
{\cal H}_{\rm com}=-\frac{c_s^2\Pi_{\calR}^2}{4a^2}-k^2a^2{\calR}^2\,.
\ea

Now we are in a position to discuss about the stability of the system. The mimetic field equation is given by Eq. (\ref{mimetic-eq-R}) with the sound speed given by Eq. (\ref{cs-def}). In order to avoid the gradient instability we require $c_{s}^{2}>0$. Now with $c_{s}^{2}>0$ we see from the 
Hamiltonian (\ref{Hamiltonian-com-conformal}) that the kinetic energy has the wrong sign, i.e. it has a ghost. In addition, independent of the sign of $c_{s}^{2}$, the gradient term in the Hamiltonian (\ref{Hamiltonian-com-conformal})  has the wrong sign.  In conclusion, for each Fourier mode the Hamiltonian is not bounded from the below since both the kinetic term and the gradient term have the wrong signs. These conclusions were also reached in \cite{Ijjas:2016pad} for the particular case of $f(\chi)  \propto \chi^{2}$ and now our analysis generalizes it for arbitrary higher derivative function $f(\chi)$. This conclusion is also consistent with the results in \cite{Ramazanov:2016xhp} who have shown that the mimetic scenario is equivalent to the IR limit of the projectable Horava-Lifshitz gravity which has the ghost instability.  
 
One natural question is how serious the above mentioned instabilities are. These issues were studied in details in  \cite{Ramazanov:2016xhp} for the IR limit of the projectable Horava-Lifshitz gravity which, as mentioned before,   is equivalent to mimetic setup. To answer this question one may wonder what the UV completion of the mimetic setup is. For instance, if it is similar to Lorentz violating scenario such as studied in  \cite{Ramazanov:2016xhp}, then the model contains higher spatial derivative operators 
inherent to the UV theory. As a result  higher gradient terms are naturally turned on at sufficiently high spatial momentum scales. This can help to eliminate the gradient instability should one start with $c_{s}^{2}<0$. However, the drawback is that for very small $c_{s}$, required to ensure that the gradient instability does not propagate  on a time scale smaller than the age of the Universe,  the strong coupling becomes too low invalidating the perturbative approach. In contrast, one may look for the branch in which $c_{s}^{2}>0$ but with  the ghost instability. If one  assumes that the scale of strong coupling and the UV scale of the theory are the same, it is concluded in \cite{Ramazanov:2016xhp} that one requires $\sqrt \gamma \ge 10  $ MeV  in the IR limit of the projectable Horava-Lifshitz gravity with $f = \gamma (\Box \phi)^{2}$ for the model to be phenomenologically viable.  We expect that these conclusions can be extended to our setup as the mimetic model is equivalent to the IR limit of the projectable Horava-Lifshitz gravity. This requires detail investigation which is beyond the scope of the current work. 
 
 Since the theory contains higher derivative terms via $f(\chi)$, one may wonder if the ghost in (\ref{action-R}) is the so-called Ostrogradsky ghost \cite{Woodard:2015zca}. 
More precisely, the Ostrogradsky ghost arises when the higher derivative terms increase the number of degrees of freedom for the system under consideration. This is however not the case for the mimetic scenario. Indeed, as discussed in \cite{Chamseddine:2014vna}, the mimetic constraint (\ref{mimetic-const.}) prevents the propagation of an extra Ostrogradsky-like degree of freedom.
This fact can be also seen from the Hamiltonian (\ref{Hamiltonian-com-conformal}) which does not contain the well-known linear term in momentum (such as the term $P_1 Q_2$  in Eq. (\ref{H-Nb}) in next section),   which arises in theories that suffer from the Ostrogradsky ghost. Therefore, we conclude that the instabilities associated with (\ref{action-R}) are usual dynamical instabilities and are  independent of the Ostrogradsky ghost.

Now we comment for the simple case $f=0$ so $c_{s}=0$. Following the same steps as before we conclude $\alpha=0$ while the constraint to eliminate $ \beta$ simply yields $\dot \calR=0$. This indicates that $\calR$ is not a propagating degree of freedom. This is in line with the fact that $c_{s}=0$ and  there is no notion of quantum wave sourcing the mimetic field perturbations. The perturbations in the mimetic  fluid with no pressure generate caustic singularities in dark matter perturbations  so the mimetic setup with  $c_{s}=0$ may not be appealing, however see \cite{DeFelice:2015moy, Gumrukcuoglu:2016jbh} and  \cite{Babichev:2016jzg, Babichev:2017lrx} where it is argued that this may not be a real problem. 

The comoving gauge has the advantage that one kills the $\delta \phi$ fluctuations so after imposing the 
mimetic constraint  the action  is manifestly  free from higher derivative terms. However, it is instructive to look at the perturbations in Newtonian gauge  which is also the gauge employed in \cite{Chamseddine:2014vna}.

\section{Newtonian gauge }
\label{Newton-sec}

Here we perform the perturbation analysis in Newtonian gauge. As we shall see, this gauge has the advantage that one can obtain a different insight into the origins of instabilities in mimetic matter scenario.
In addition, this gauge enables us to compare our results with those of \cite{Chamseddine:2014vna}. We calculate  the quadratic action, the step which was not performed in  \cite{Chamseddine:2014vna}, which  is necessary to investigate the stability of the system. 

As usual, in this gauge the metric perturbation is given by
\ba
ds^{2 } = - ( 1 + 2 \Phi) dt^{2 } + a^{2} ( 1- 2 \Psi) \delta_{i j} dx^{i } dx^{j} \, ,
 \ea
in which $\Phi$ and $\Psi$ are the Bardeen potentials. Happily, the mimetic stress energy tensor Eq. (\ref{Tmu-nu}) has no spatial off-diagonal term so we conclude $\Psi= \Phi$ and in the rest of analysis we work with $\Phi$. 

Imposing the mimetic constraint Eq. (\ref{mimetic-const.}) we obtain
\ba
\label{Newton-const.}
\Phi = \delta \dot \phi \, .
\ea
In addition, from the $0i$ component of Einstein equation we obtain
\ba
\label{0i-const.}
-2 \left( \dot \Phi + H \Phi  \right) =  ( 2 \bar \lambda + 3 
\dot H f_{\chi \chi}) \delta \phi -  f_{\chi \chi} 
\delta \chi ^{(1)}  \, .
\ea
in which the perturbation $\delta \chi ^{(1)}$ to linear order  is given by
\ba
\label{chi-1}
\delta \chi ^{(1)}=  -\delta \ddot \phi - 3 H \delta \dot \phi + 
6 H \Phi + 4 \dot \Phi + \frac{\partial ^2\delta \phi }{a^2}\, .
\ea
Combining the constraint equations (\ref{Newton-const.}) and  (\ref{0i-const.}) and using the value of
$\bar \lambda $ given in Eq. (\ref{lambda-eq0}), we obtain the $\delta \phi$ equation of motion 
\ba
\label{mimetic-eom}
 \delta  \ddot \phi + H \delta \dot \phi + \dot H \delta \phi - \frac{c_{s}^{2}}{a^{2}} \partial^{2 } \delta \phi 
 =0 \, ,
\ea
in which $c_{s}^{2}$ is the sound speed of $\delta \phi$ perturbation defined in Eq. (\ref{cs-def}). 

The  point to emphasis is that the equation of motion (\ref{mimetic-eom}) is obtained purely from the constraint equations  (\ref{Newton-const.}) and  (\ref{0i-const.}) without resorting to the remaining $00$ or $ii$ components of Einstein equations.  One can check that  the other components of  Einstein equations are satisfied if one implements the equation of motion  Eq. (\ref{mimetic-eom}). 

Logically, there is nothing wrong in obtaining the field equations purely from the constraint equations. However, to investigate the stability of the theory, we still need to calculate the quadratic action and the Hamiltonian and then to check if the theory is healthy. This last step was not performed in \cite{Chamseddine:2014vna}. Indeed, in \cite{Chamseddine:2014vna} the field equation of motion is obtained as outlined above without calculating the quadratic action. The point we emphasis is that a well-behaved field equation by itself may not guarantee the underlying theory is safe. For example, if one multiplies the action by an overall minus sign 
the field equation does not change. But a Lagrangian with a wrong overall sign yields to a sick Hamiltonian.

With these discussions in mind, we proceed to calculate the quadratic action in Newtonian gauge, $S_{N}^{(2)}$.  Using the relation $\sqrt{-g} = a^{3} ( 1- 2 \Phi - 2 \Phi^{2})$,  we group different parts of 
the action as follows
\ba
S_{N}^{(2)} = \int d^{4} x \left[ L_{EH}^{(2)}+ a^3\left(L_{M}^{(2)} -2 \Phi  L_{M}^{(1)}-2 \Phi^{2} 
L_{M}^{(0)}\right)  \right] \, ,
\ea
in which $L_{EH}^{(2)}$ represents the contributions from the Einstein-Hilbert term while $ L_{M}^{(i)} $ indicates the contributions of mimetic matter, i.e. the contributions of the last three terms in the original action  (\ref{action0}).  

Calculating each term separately, we have 
\ba
\label{L-gr}
L_{EH}^{(2)} = 6 a^{3} \dot H \Phi^{2} - 3 a^{3} \dot \Phi^{2}  - a (\partial \Phi)^{2} \, ,
\ea
and
\ba
\label{LM0}
L_{M}^{(0)} = f(\bar \chi) - V(\bar \phi) \, ,
\ea
in which $\bar \chi$ and $\bar \phi$ indicate the background values of $\chi$ and $\phi$. In addition,
\ba
L_{M}^{(1)} = \frac{2 f_{\chi }}{f_{\chi \chi}} ( \delta  \ddot \phi + H \delta \dot \phi + \dot H \delta \phi ) - V' \delta \phi - 3 f_{\chi} \dot H \delta \phi \, ,
\ea
and
\ba
L_{M}^{(2)} = f_{\chi} \delta \chi ^{(2)} +\frac{f_{\chi \chi}}{2} (\delta \chi^{(1)} )^{2}
-\frac{1}{2} V'' \delta \phi^{2} + \bar \lambda \big( - \delta \dot \phi^{2} +
\frac{1}{a^{2} } (\partial \delta \phi)^{2}\big) \, ,
\ea
in which $\delta \chi^{(1)} $ and $\delta \chi^{(2)} $ represent the first and second order perturbations in 
$\chi$. The first order perturbation $\delta \chi ^{(1)} $ is already presented in Eq. (\ref{chi-1}) while the second order perturbation is given by 
\ba
\delta \chi^{(2)} = 2 \delta \dot \phi  \big( \delta  \ddot \phi- 3 H \delta \dot \phi + \frac{1}{a^{2}} \partial^{2} \delta \phi  \big) \, .
\ea

Combining the four different contributions in $S_{N}^{(2)}$, performing some integrations by parts, and going to conformal time, the quadratic action is obtained to be
\ba
\label{action-Na}
S_{N}^{(2)}= \int d^3 x d \tau \Bigg[& \frac{3}{2} ( 3 f_{\chi \chi} -2)   \left(  \delta \phi'' - \epsilon a^2 H^2  \delta \phi \right)^2  - \epsilon a^2 H^2 (\partial \delta \phi)^2 - (\partial  \delta  \phi')^2  \nonumber\\
&+ 3  f_{\chi \chi} \left(  \delta \phi'' - \epsilon a^2 H^2 \delta \phi \right) \partial^2 \delta \phi
+ \frac{1}{2 } f_{\chi \chi} \left( \partial^2 \delta \phi \right)^2 \Bigg] \, ,
\ea
in which a prime indicates the derivative with respect to conformal time. 
We comment that in obtaining the above action we did not use the equation of motion Eq. (\ref{mimetic-eom})  and only used the constraint $\Phi= \delta \dot \phi$ to eliminate $\Phi$. 

At this form the action (\ref{action-Na}) looks problematic as it has higher derivative terms so naively it leads to Ostrogradsky ghost \cite{Woodard:2015zca}. Indeed, varying the action (\ref{action-Na}) with respect to $\delta \phi$ yields the equation of motion as
\ba
\label{full-eom}
 \Omega'' -  \epsilon a^2 H^2 \,  \Omega + \frac{ \partial^{2} \Omega}{3}   =0  \, ,
\ea
in which for simplification, we have defined 
\ba
 \Omega \equiv  3 ( 3 f_{\chi \chi} -2 ) (\delta  \phi''  -  \epsilon a^2 H^2)  +  3 f_{\chi \chi }\partial^2  \delta \phi. 
 \ea
 The advantage of defining $\Omega$ as above is that the mimetic field equation (\ref{mimetic-eom}) simply translates into $\Omega =0$. 
 
The higher derivative structure of the field equation in Eq. (\ref{full-eom}) is evident. However, this by itself does not indicate the theory is sick. The reason is that  we have not yet imposed the constraint equation (\ref{mimetic-eom}) which is $\Omega=0$. One can obviously see that the mimetic equation $\Omega=0$  
is satisfied by the higher derivative equation (\ref{full-eom}). This is a consequence of the mimetic constraint Eq. (\ref{mimetic-const.}) which protects the theory from the dangerous higher derivative terms. Therefore, the evolution of the system is given by the constraint field equation $\Omega=0$ which requires only two initial data on the initial Cauchy hypersurface.  This should be compared with the naive field equation  (\ref{full-eom}) which requires four initial data on initial Cauchy hypersurface.

In conclusion, the theory is safe under higher derivatives and it does not have Ostrogradsky ghost. For this to happen, it is crucial to  implement the
constraint $\Omega=0$ into the higher derivative action (\ref{action-Na}). 

As a consistency check, one may wonder if the action (\ref{action-Na}) is consistent with the action 
(\ref{action-R}) in comoving gauge. To verify this note that $\calR$ in comoving gauge is related to  $\delta \phi$ in Newtonian gauge via  $\calR \rightarrow  - (\delta \dot \phi + H \delta \phi)$. Then imposing the constraint equation $\Omega=0$ and after some integration by parts we obtain Eq. (\ref{action-R}). 
Note that it is crucial to use the constraint equation $\Omega=0$ in Newtonian gauge 
 to check the equivalence of the two actions since in obtaining the action (\ref{action-R})  in comoving gauge  we have used all constraint equations as given in Eq. (\ref{solution1}).

\subsection{Hamiltonian analysis in Newtonian gauge}
\label{Hamiltonian-sec}

Having obtained the action, here we calculate the Hamiltonian of the system in Newtonian gauge to identify the instabilities as viewed in this gauge. 

As usual, we have to obtain the Hamiltonian from the Lagrangian by constructing  the conjugate momenta. Our starting Lagrangian is Eq. (\ref{action-Na})  subject to constraint of equation of motion, $\Omega=0$. One direct  strategy is to simply insert the constraint $\Omega=0$ into the action (\ref{action-Na}) to get rid of the higher derivative terms. Going to Fourier space, the Lagrangian after imposing the mimetic constraint is given by 
\ba
\label{L-Fourier-N}
{\cal L}_{\rm N}\Big{|}_{\Omega=0} = - k^2 \delta\phi'^2 + k^2 (c_s^2 k^2 - \epsilon a^2 H^2) \delta \phi^2 \, .
\ea
From the above Lagrangian the  Hamiltonian is obtained to be 
\ba
\label{H-Fourier-N}
{\cal H}_{\rm N}\Big{|}_{\Omega=0}  = -\frac{\Pi_{\delta\phi}^2}{4 k^2} - 
k^2 ( c_s^2 k^2 - \epsilon a^2 H^2) \delta\phi^2 \, ,
\ea
in which $\Pi_{\delta\phi} =  \partial {\cal L}/\partial \delta\phi'=  -2 k^2 
\delta\phi'$ is the conjugate momentum.

In the above analysis, we have obtained the Hamiltonian after imposing the constraint $\Omega=0$ directly into Lagrangian.  It is also instructive to construct the Hamiltonian instead by imposing the constraint $\Omega=0$ in full higher dimensional phase space. For this purpose, note that the phase space associated with the action (\ref{action-Na}) is four-dimensional spanned by the canonical variables $(P_1, Q_1, P_2, Q_2)$ in which $P_i$ and $Q_i$ are constructed following Ostrogradsky method   where \cite{Woodard:2015zca}
\ba
Q_{1 } \equiv\delta \phi, \quad  Q_{2} \equiv \delta \phi'    \, ,
\ea
while the conjugate momenta are given by
\ba
P_{1 }= \frac{\partial L}{\partial \delta \phi'} - \frac{d}{d \tau} \frac{\partial L}{\partial \delta \phi''}
\quad, \quad  P_{2 }= \frac{\partial L}{\partial \delta \phi''} \, .
\ea 
In particular, from action (\ref{action-Na}) the conjugate momentum $P_2$ is  obtained to be 
\ba
\label{P2-Omega}
P_2 = 3 ( 3 f_{\chi \chi} - 2) \left( \delta \phi'' + ( c_s^2 k^2 - \epsilon a^2 H^2) \delta \phi
\right)=    \Omega  \, .
\ea
Interestingly we see that the constraint of equation of motion $\Omega=0$ in phase space is translated into $P_2=0$. We should implement this constraint when constructing the Hamiltonian. 

The Hamiltonian is constructed as usual via ${\cal H}_{N} (P_1, Q_1, P_2, Q_2) = P_1 Q_1' + P_2 Q_2' - {\cal L}$ yielding  
\ba\label{H-Nb}
{\cal H}_{\rm N} = P_{1} Q_{2}- \frac{1+ 3c_s^2 }{12 } P_{2}^{2} - ( c_s^2 k^2 - 
\epsilon a^{2 }  H^{2} ) Q_1 P_2 + k^2 Q_2^2 - k^2 ( c_s^2 k^2 -  
\epsilon a^{2 }  H^{2} ) Q_1^2  \, .
\ea
The first  term which is linear in $P_1$ is the hallmark of the Ostrogradsky ghost. Indeed, if the Hamiltonian was just as in Eq. (\ref{H-Nb}), then we would have had the Ostrogradsky ghost. However, we still have to implement the constraint $P_2=0$ in our Hamiltonian. Furthermore,  the  evolution of $P_2$ is governed by  $\dot P_2 = -\partial {\cal H}_{\rm N}/\partial Q_2$. The consistency of the constraint $P_2=0$  requires $\dot P_2 =0$. This in turn  imposes an additional constraint 
\ba
\label{P2-eq}
P_1+ 2 k^2 Q_2=0 \, .
\ea
Now, eliminating $Q_2$ in favor of $P_1$ from Eq. (\ref{P2-eq}) and imposing $P_2=0$, the total Hamiltonian from Eq. (\ref{H-Nb}) reduces to the following form 
\ba\label{H-red} 
{\cal H}_{\rm N} = -\frac{P_1^2}{4 k^2} - k^2 ( c_s^2 k^2 - \epsilon a^2 H^2 ) Q_1^2  \, .
\ea
As expected, the four-dimensional  phase space is reduced to two-dimensional phase space $(Q_1, P_1)$ which is a consequence of the constraint $P_2=0$. This is another realization that the true equation of motion is only second order and not a fourth order differential equation. In addition, we see that the  reduced Hamiltonian obtained above is the same as one obtains from the simple replacement of the constraint in Lagrangian given in Eq. (\ref{H-Fourier-N}).

One may ask if the question of instabilities and the form of Hamiltonian obtained above are consistent with those obtained in comoving gauge. To clarify this point, we will show that the two  Hamiltonian functions (\ref{H-red}) (or equivalently (\ref{H-Fourier-N}))  and (\ref{Hamiltonian-com-conformal}) in Newtonian and comoving  gauges respectively are related to each other through a canonical 
transformation and therefore they describe the same physical system in different coordinates. The coordinates  $\calR$ in comoving gauge and $\delta \phi$ in Newtonian gauge are related via  $\calR= -a^{-1}\delta\phi'-H\delta\phi$ which leads us to the following  canonical  transformation  between the
coordinates $(Q_{1}, P_{1})$ and $({\calR}, \Pi_{\calR})$ 
\ba
\label{CT-com-New}
\calR=\frac{P_1}{2ak^2}-H Q_1\,,\hspace{1cm}\Pi_\calR=-2ak^2Q_1\,,
\ea
where $\Pi_\calR$ is constructed  such that $\{\calR,\Pi_{\calR}\}=1$ in order  to have a consistent canonical transformation. 

Applying the canonical transformation (\ref{CT-com-New}) into  the invariant Lagrangian density ${\cal L}_{\rm N}=Q'_1P_1- {\cal H}_{\rm N}(Q_1,P_1)$ and then using an appropriate integration 
by part, one  finds the following expression for the  associated generating function   
\ba\label{GF-com-New}
\frac{\partial {\cal F}}{\partial \tau}=(H^2+a^{-1}H')\frac{\Pi_{\calR}^2}{ 4k^2}-aH\, \calR \Pi_{\calR}\,.
\ea
The transformed Hamiltonian ${\cal H}_{\rm N}(\calR,\Pi_{\calR})+
\partial {\cal F}/\partial\tau$ then immediately leads to the Hamiltonian 
(\ref{Hamiltonian-com-conformal}) in comoving gauge. Therefore,
the results in both gauges are consistent. 

In conclusion, in both comoving and Newtonian gauges we have shown that the Hamiltonian in the mimetic matter scenario is not bounded from the below such that both the kinetic term and the gradient term have the wrong signs. Although the action in Newtonian gauge contains the Ostrogradsky higher derivative terms like ${\delta\phi''}^2$, we have shown that the mimetic constraint ensures  the theory to be free of Ostrogradsky-like ghost. Therefore, the instabilities in mimetic scenario are usual dynamical instabilities which are not related to the Ostrogradsky ghost.

In both gauges we have imposed the mimetic constraint equation and the constraint from the $0i$ component of Einstein equation during the construction of quadratic action. This may raise  some concerns that we may have lost some hidden properties of the model or inflicted unwanted properties to the setup. In order to clarify this question, in next section we perform the full Hamiltonian analysis with no constraints imposed by hand. Of course, on the physical ground we do not expect any difference to appear compared to results we have obtained so far. Therefore, the reader who is convinced by now on the existence of instabilities may skip the analysis of next section. 

\section{Full Hamiltonian analysis}
\label{Hamiltonian-secf}

In this section we perform the full Hamiltonian analysis  without imposing the mimetic constraint or other constraints  at the perturbation level  by hand.
Our only assumption is to fix the gauge,  working in comoving gauge which is easier to work with. 
The Hamiltonian analysis for the mimetic scenario were performed in \cite{Chaichian:2014qba, Malaeb:2014vua, Barvinsky:2013mea}. In particular the analysis of  \cite{Chaichian:2014qba, Malaeb:2014vua} are mostly focused  on the formalism of Hamiltonian approach.  Here, we focus less on formalism but write down our analysis specifically for the relevant cosmological perturbations which are necessary in studying  the stability of the scenario. 

Our starting point is the full action in comoving gauge, Eq. (\ref{action}).   Substituting the metric perturbations  (\ref{hij}) and (\ref{ADM-scalar-pert})  in action (\ref{action}) and going to Fourier space, we obtain the following 
quadratic Lagrangian (see the Appendix for details of derivation)
\ba
\label{S2-comoving}
\dfrac{{\cal L}_{\rm com}^{\rm tot}}{a^3}&=&-3\dot{\calR}^2- (18H\calR
-6H\alpha+2k^2\beta)\dot{\calR}-\left(\frac{9}{2}\dot{H}
+27H^2-\frac{k^2}{a^2}\right)\calR^2-3H^2\alpha^2\\
\nonumber
&+&2k^2H\alpha\beta-\left(3\dot{H}-\frac{2k^2}{a^2}
\right)\alpha\calR+\left[(\alpha+3\calR)+(\alpha-3\calR)
\dot{\phi}^2\right] \lambda^{(1)}-(\dot{\phi}^2+1) 
\lambda^{(2)}\\
\nonumber
&-&\dot{\phi}f_{\chi}\left[3H(\alpha-6\calR)\alpha-3(
\alpha-3\calR)\dot{\calR}+(2\alpha-3\calR)\dot{\alpha}
\right]-\frac{9}{2}H f_{\chi}(2\alpha+3\calR)\calR\\
\nonumber
&+&\dot{\phi}^2\left[f_{\chi\chi}\left(6H\alpha+
\dot{\alpha}-3\dot{\calR}-k^2\beta\right)^2-\dot{H}
(1-3f_{\chi\chi})(2\alpha^2+9\calR^2-6\alpha\calR)
\right]\\
\nonumber
&-&\ddot{\phi}\left[f_{\chi}(\alpha-6\calR)-2
\dot{\phi} f_{\chi\chi}\left(6H\alpha+\dot{\alpha}
-3\dot{\calR}-k^2\beta\right)\right]\alpha+2
\ddot{\phi}^2 f_{\chi\chi} \alpha^2\,.
\ea
As mentioned at the start of this section, we do not impose the mimetic constraint (\ref{mimetic-const.})
by hand. Therefore, we allow the Lagrange multiplier $\lambda$ to vary with $\lambda^{(1)}$ and $\lambda^{(2)}$ respectively  representing  the first and second order perturbations in $\lambda$. 

Our goal is to construct the Hamiltonian associated with the quadratic action (\ref{S2-comoving}). There are five coordinate variables $Q_i=(\alpha,\beta, \calR,\lambda^{(1)},\lambda^{(2)})$ so the  phase space is ten-dimensional spanned by  
$Q_i$ and the corresponding conjugate momentum 
$P_i=\partial {{\cal L}_{\rm com}^{\rm tot}}/\partial{\dot{Q}_i}$.  
The standard Poisson algebra $\{Q_i,P_j\}=\delta_{ij}$ holds between the 
canonical coordinates and momenta. Note that since we work in Fourier space, the variables $Q_i$ and $P_i$ are only functions of time and for 
simplicity we drop their dependence on $k$, i.e, we denote $Q_i(t,k) $ 
simply by $Q_i(t)$ and so on. 

Looking at the Lagrangian (\ref{S2-comoving}) we see that $P_2= P_4= P_5=0$, i.e. the
coordinates $\beta, \lambda^{(1)}$ and $\lambda^{(2)}$ have no time derivatives. These 
means we have three primary constraints 
\ba
\label{Primary-C}
\Upsilon_1\equiv P_2=0\,,\hspace{.5cm}
\Xi_1\equiv P_4=0\,,\hspace{.5cm}
\Theta_1\equiv P_5=0\,,\hspace{.5cm}
\ea
while for $P_1$ and $P_3$ we have 
\ba\label{Cmomenta}
P_1=P_1(Q_1,Q_2,Q_3,\dot{Q}_1,\dot{Q}_3),\hspace{.5cm}
P_3=P_3(Q_1,Q_2,Q_3,\dot{Q}_1,\dot{Q}_3) \, .
\ea

Solving (\ref{Cmomenta}) for $\dot{Q}_1$ and $\dot{Q}_3$ in terms of the phase space 
variables $(Q_i,P_i)$ and taking into account the primary 
constraints (\ref{Primary-C}), we construct the total 
Hamiltonian function from the standard definition,  see for example \cite{Chen:2012au, Motohashi:2016ftl, Deffayet:2015qwa}, as follows 
\ba
\label{T-Hamiltonian} 
{\cal H}_{\rm com}^{tot}=\dot{Q}_iP_i-L+u_1\Upsilon_1+v_1\Xi_1+w_1\Theta_1\,,
\ea
where $u_1$, $v_1$, and $w_1$ are Lagrange multipliers  (auxiliary fields) which enforce the constraints (\ref{Primary-C}).  The form of the total Hamiltonian function (\ref{T-Hamiltonian})  is complicated which is  given in Appendix in Eq. (\ref{T-Hamiltonian1}). 

In order to identify the secondary constraints, we should consider the  time evolution of the primary constraints (\ref{Primary-C})  which amount to check for consistency relations.  We have seen a simple form of this consistency condition in deriving Eq. (\ref{P2-eq}) in 
previous section. Starting with $\Theta_1$, we obtain the following constraint  
\ba\label{Theta1dot}
\Theta_2 \equiv \{\Theta_1, {\cal H}_{\rm com}^{\rm tot}\}=
a^3(\dot{\phi}^2-1)\approx 0\,,
\ea
in which, following \cite{Chen:2012au}, the notation $\approx 0$ stands for constraint equations. From the above condition we immediately obtain the  expected  result
\ba
\label{phidot}
\dot{\phi}=1\,,
\ea
which is nothing but the mimetic matter constraint.  In addition, relation (\ref{Theta1dot}) shows 
that $\Theta_1=0$ is a first class constraint. 

The time evolution of  $\Xi_1=0$, however, leads to another constraint 
\ba\label{Xi1dot}
\Xi_2 \equiv \{\Xi_1,{\cal H}_{\rm com}^{\rm tot}\}=a^3\left(Q_1(\dot{\phi}^2+1)
-3Q_3(\dot{\phi}^2-1)\right)
\approx 0\,.
\ea
Imposing condition (\ref{Theta1dot}), the above constraint can  be solved for $Q_1$ which gives
\ba\label{Q1}
Q_1=0\,,
\ea
in agreement with the previous result $\alpha=0$ in (\ref{solution1}). The next consistency relation
generates another new constraint 
\ba\label{Xi2dot}
\Xi_3 \equiv \{\Xi_2, {\cal H}_{\rm com}^{\rm tot}\}=
\Xi_3(Q_1,Q_2,Q_3,P_1,P_3)\approx 0\,.
\ea
Requiring this constraint to be time independent,
gives
\ba
\label{Xi3dot}
\Xi_4 \equiv \{\Xi_3, {\cal H}_{\rm com}^{\rm tot}\}=
\Xi_4(u_1;Q_1,Q_2,Q_3,Q_4,P_1,P_3)\approx 0\,,
\ea
which determines the Lagrange multiplier $u_1$ in terms of phase space variables and so the chain of 
constraints for primary constraint $\Xi_1$  terminates here.

For the time evolution of the next primary constraint  $\Upsilon_1=0$, we obtain  the following new constraint 
\ba
\label{Upsilon1dot}
\Upsilon_2 \equiv \{\Upsilon_1, {\cal H}_{\rm com}^{\rm tot}\}=
\Upsilon_2(Q_1,Q_2,Q_3,P_3)\approx 0\,.
\ea
The time independence of this new constraint gives
\ba
\label{Upsilon2dot}
\Upsilon_3=\{\Upsilon_2, {\cal H}_{\rm com}^{\rm tot}\}=
\Upsilon_3(u_1;Q_1,Q_2,Q_3,Q_4,P_1,P_3)\approx 0\,,
\ea
which determines the Lagrange multiplier $u_1$ and
so the chain of constraints for the primary constraint
$\Upsilon_1$ terminates here. 

The above considerations show that the three primary 
constraints (\ref{Primary-C}) generate three secondary 
constraints (\ref{Xi1dot}), (\ref{Xi2dot}), and 
(\ref{Upsilon1dot}). Therefore, the system under consideration
admits $6$ constraints $C_i=(\Upsilon_1,\Upsilon_2,\Xi_2,\Xi_3,
\Theta_1,\Xi_1)$. The associated Dirac matrix with entries
$D_{ij}=\{C_i,C_j\}$ turns out to be
\begin{eqnarray}\label{DiracMatrix}
D= a^3\left(
\begin{matrix}
	0 & -\frac{2}{3}k^4 & 0 & 0 & 0 & 0 \\ 
	\frac{2}{3}k^4& 0 & 0 & 0 & 0 & 0 \\ 
	0 & 0 & 0 & 2(-3+f_{\chi\chi}^{-1}) & 0 & 0 \\ 
	0 & 0 & 2(-3+f_{\chi\chi}^{-1}) & 0 & 0 & 0 \\ 
	0 & 0 & 0 & 0 & 0 & 0 \\ 
	0 & 0 & 0 & 0 & 0 & 0
\end{matrix} 
\right),
\end{eqnarray}
where we used relation (\ref{phidot}).

The rank of Dirac matrix (\ref{DiracMatrix}) determines the number of 
second class constraints which is four and the remaining  two are 
first class. In this respect, the physical number of degrees of freedom 
is $(10-2\times{2}-4)/2=1$ which is nothing but the curvature 
perturbation $\calR$ in comoving gauge. Therefore, we can determine 
all the phase space variables $(Q_i,P_i)$ in terms of two phase space 
variables $(Q_3,P_3)$ which denote the curvature perturbation and its 
conjugate momentum.

In order to do this, we first substitute (\ref{Q1}) in 
(\ref{Upsilon2dot}) and then solve for $Q_2$ which gives 
\ba
\label{Q2}
Q_2=-\frac{P_3}{2k^2a^3}+\frac{9(2H+f_{\chi})}{2k^2}\,Q_3\,.
\ea
We have checked that the above value for $Q_2 =\beta$, agrees with the result obtained in
Eq. (\ref{solution1}). 

Finally, substituting Eqs. (\ref{Q2}) and  (\ref{Q1}) into the constraint (\ref{Xi2dot}), and imposing condition (\ref{phidot}), we obtain  the following solution for the  conjugate momentum $P_1$
\ba\label{P1}
P_1=c_s^2P_3-6 c_s^2 a^3(3H+f_{\chi}f_{\chi\chi})\,Q_3\,,
\ea
where we have substituted speed of sound from (\ref{cs-def}).

In order to determine $Q_4$ in terms of $(Q_3,P_3)$, we solve  $u_1$ from the relations (\ref{Xi3dot}) and (\ref{Upsilon2dot}). Equating the two results for $u_{1}$ and substituting from (\ref{phidot}), (\ref{Q1}), (\ref{Q2}) and (\ref{P1}), gives
\ba
\label{Q4}
Q_4=-\frac{c_s^2 H}{a^3} (3-f_{\chi\chi}^{-1}) P_3
+\frac{2c_s^2k^2}{a^2}\left[(2-f_{\chi\chi}^{-1})-9a^2H
(3(3H+f_{\chi})-2Hf_{\chi\chi}^{-1})\right] Q_3\,.
\ea

Now, all the phase space variables are calculated as functions of   $(Q_3,P_3)$ so the  reduced 
Hamiltonian is given entirely in in terms of  $(Q_3,P_3)$. Plugging the corresponding variables as functions of   $(Q_3,P_3)$ the reduced Hamiltonian is obtained to be 
\ba
\label{R-Hamiltonian}
{\cal H}_{\rm com}=A\,P_3^2+B\,Q_3P_3+C\,Q_3^2\,,
\ea
where the coefficients $A$, $B$, and $C$ are defined as
\ba\label{R-HamiltonianAB}
A=-\frac{c_s^2}{4a^3}\, ,\hspace{1cm}
B=-\frac{9}{2} c_s^2 (2H+f_{\chi})\,,
\ea
and 
\ba
\label{R-HamiltonianC}
C=-ak^2+\frac{9}{4}a^3 c_s^2\left[2c_s^{-4}\dot{H}
f_{\chi\chi}-9f_{\chi}^2+24H^2 f_{\chi\chi}^{-1}\Big(1+
6Hf_{\chi}(2-9f_{\chi})-3f_{\chi\chi}\Big)\right]\,.
\ea

The Hamiltonian function (\ref{R-Hamiltonian}) can be  transformed into the diagonal form via the following canonical  transformation
\ba\label{CT}
\calR=Q_3 \, ,\hspace{1cm}\Pi_{\calR}=P_3+\frac{B}{2A}Q_3\,.
\ea
Under the above canonical transformation, the Hamiltonian function transforms as  
$  {\cal H}_{\rm com} \rightarrow {\cal H}_{\rm com}(\calR,\Pi_{\calR})+\partial  {\cal G}/{\partial t}$ in which ${\cal G}$ is the associated generating 
function that solves the equation $P_3={\partial {\cal G}}/{ \partial Q_3}$. Using Eq.  (\ref{CT}), we obtain the following solution  for the generating function
\ba\label{GF}
{\cal G}(t;\calR,\Pi_{\calR})=\calR \Pi_{\calR}-\frac{B}{4A}\calR^2\,,
\ea
up to an arbitrary additive function of time alone which does  not affect the 
action. The Hamiltonian function then  transforms as  
\ba
\label{H-transformed}
 {\cal H}_{\rm com} \rightarrow    {\cal H}_{\rm com} -\frac{1}{4}\left(\frac{\dot{B}}{A}
-\frac{B\dot{A}}{A^2}\right)\,\calR^2 \, .
\ea
Now substituting  Eqs.  (\ref{R-HamiltonianAB}) and  (\ref{R-HamiltonianC})  in Eq. (\ref{H-transformed}) gives our final result for the reduced Hamiltonian 
\ba
\label{R-Hamiltonian-diag}
{\cal H}_{\rm com}=-\frac{c_s^2\Pi_{\calR}^2}{4a^3}-a k^2 \calR^2\,.
\ea
After going to conformal time, we see that  the above Hamiltonian correctly coincides with 
the Hamiltonian previously obtained in Eq. (\ref{Hamiltonian-com-conformal}). The conclusion on the instabilities of  the mimetic scenario is the same as before.

\section{Discussions }
\label{summary-sec}

In this paper we have revisited cosmological perturbations in mimetic scenario with a  general higher derivative function $f(\chi)$. The higher derivative terms are needed to generate gradient terms for the mimetic perturbations to prevent the formation of caustic singularities in dark matter perturbations. In addition, the gradient terms make  the notion of quantum wave associated with the mimetic field perturbations well-defined.

We have confirmed that the mimetic constraint Eq. (\ref{mimetic-const.}) prevents the appearance of higher derivative ghosts. This is most easily seen in comoving gauge. In Newtonian gauge, however, the mimetic field equation appears as a constraint equation after imposing the $0i$ and the 
mimetic constraint  (\ref{mimetic-const.}). While the resulting action in Newtonian gauge has a higher derivative form, but it reduces to a well-behaved form after imposing the mimetic equation of motion as a constraint.

In both gauges we have seen that both the kinetic term and the gradient term have the wrong signs
so the Hamiltonian is not bounded from the below. This indicates the model in its simple realization with no UV completion is pathological.  If one assumes $c_{s}^{2}>0$, then the system has no gradient instability but it has ghost instability.  We emphasis that these instabilities are not associated with the higher derivative ghosts which may naively come to mind.  Finally, we have shown that the Hamiltonians in these two gauges are related to each other via a canonical transformation so the physical conclusion on the existence of instabilities are consistent with each other in both gauges.

The appearance of instabilities in the current realization of mimetic setup is unavoidable. It is an interesting question to see if one can avoid instabilities in an extension of the mimetic setup. We hope to come back to this question elsewhere.  \\


{\bf Acknowledgments:}  We would like to thank  Frederico Arroja, 
Javad T. Firouzjaee, Sabino Matarrese and  Shinji Mukohyama for useful discussions and correspondences.

\vspace{0.7cm}

\appendix

\section{Total Lagrangian and Hamiltonian}

Here we briefly outline the steps of analysis leading to quadratic action  Eq. (\ref{S2-comoving}).

The full action in comoving gauge  is given in Eq. (\ref{action}), which  we decompose it into various combinations as follows 
\ba
S_{\rm com}^{(2)} = \int d^{4}x\, a^3
\left(L_{EH}^{(2)} + { L}_M^{(2)} + (\alpha+3 \calR) { L}_M^{(1)}
+ (3 \alpha \calR +\dfrac{9\calR^2 }{2}) { L}_M^{(0)}
\right)\,,
\ea
in which $L_{EH}^{(2)}$ represents the contribution of the Einstein Hilbert term given by 
\ba
L_{EH}^{(2)}
&=& -3\dot{\calR}^2-18H \calR\dot{\calR}-\dfrac{27}{2} H^2 \calR^2+6H \alpha\dot{\calR}+ 9 H^2 \alpha \calR- 3 H^2 
\alpha^2
\\ \nonumber 
&-&\frac{1}{a^2}\Big(\left(\partial\calR\right)^2+2(\alpha+\calR)
\partial^2\calR\Big)-2H (\alpha-3\calR)\partial^2\beta+
6H\partial_i\calR \partial_i\beta+2\dot{\calR}\partial^2\beta
\\ \nonumber
&+&\left(\partial_i\partial_j\beta\partial_i\partial_j\beta
-\left(\partial^2\beta\right)^2\right)\, .
\ea
The rest are the contribution of the mimetic matter sector which are grouped in three different terms as follows  
\ba\label{LM0-com}
L_{M}^{(0)} =  f(\bar{\chi})+\dot{H}(\dot{\phi}^2-1)(3f_{\chi\chi}-1)\,,
\ea
\ba\label{LM1-com}
L_{M}^{(1)} = f_{\chi}\left[6H \dot{\phi}\alpha+2\ddot{\phi}\alpha
+\dot{\phi}\left(\dot{\alpha}-3\dot{\calR}+\partial^2\beta\right)
\right]+2\dot{H}\dot{\phi}^2(1-3f_{\chi\chi})\alpha-
\lambda^{(1)}(\dot{\phi}^2-1)\,,
\ea
and
\ba\label{LM2-com}
L_{M}^{(2)}  
&=& \dot{\phi}\left[2f_{\chi\chi}(3H\dot{\phi}+\ddot{\phi})
\left(\dot{\alpha}-3\dot{\calR}+\partial^2\beta\right)-f_{\chi}
\left(3\dot{\alpha}-6\dot{\calR}+2\partial^2\beta\right)\right]
\alpha\\
\nonumber
&+& \left[2f_{\chi\chi}(3H\dot{\phi}+\ddot{\phi})^2
+3\dot{H}\dot{\phi}^2(3f_{\chi\chi}-1)-3\dot{\phi}^2
f_{\chi}\right]\alpha^2+\dot{\phi}(2\lambda^{(1)}
\dot{\phi}-9Hf_{\chi}\alpha)\alpha\\
\nonumber
&+& \frac{1}{2}\dot{\phi}
\left[\dot{\phi}f_{\chi\chi}\left(\dot{\alpha}-
3\dot{\calR}+\partial^2\beta\right)^2-2f_{\chi}\partial_i
(\alpha-3\calR)\partial_i\beta\right]-\lambda^{(2)}(
\dot{\phi}^2-1)\,.
\ea
Combining the above four contributions yield the quadratic action  Eq. (\ref{S2-comoving}). 

Having obtained the quadratic action we can calculate the Hamiltonian. After imposing the primary constraints 
the total Hamiltonian is given by
\ba
\label{T-Hamiltonian-b} 
{\cal H}_{\rm com}^{tot}=\dot{Q}_iP_i-L+u_1\Upsilon_1+v_1\Xi_1+w_1\Theta_1\,.
\ea
Now constructing the conjugate momenta $P_i=\partial L/\partial{\dot{Q}_i}$  from the quadratic action 
(\ref{S2-comoving}) the total Hamiltonian is obtained to be 
\ba
\label{T-Hamiltonian1}
 {\cal H}_{\rm com}^{\rm tot}  &=&  (12 a^3 \dot{\phi}^4 f_{\chi\chi} )^{-1} 
 \times \Bigg[ 3\left(2\dot{\phi}^2(1 - 3 f_{\chi\chi})+3f_{\chi\chi}\right) P_1^2-6\dot{\phi}^4f_{\chi\chi}P_1P_3-\dot{\phi}^4
f_{\chi\chi}P_3^2\\
\nonumber
&+&6a^3\dot{\phi}\left[\left(4\dot{\phi}^2+3(\dot{\phi}^3-
3\dot{\phi}^2-\dot{\phi}+2)f_{\chi\chi}\right)f_{\chi}-2
\left(3f_{\chi\chi}+\dot{\phi}^2\ddot{\phi}(2-3f_{\chi\chi})
\right)f_{\chi\chi}\right]P_1Q_1
\\ \nonumber
&-&18a^3\dot{\phi}\left[6H\dot{\phi}^3f_{\chi\chi}+\left(
2\dot{\phi}^2+3(\dot{\phi}-1)^2(\dot{\phi}+1)f_{\chi\chi}
\right)f_{\chi}\right]P_1Q_3-36Ha^3\dot{\phi}^4 
f_{\chi\chi} P_3 Q_3\\ 
\nonumber
&+&3a^6\dot{\phi}^2f_{\chi}\left[12\ddot{\phi}
f_{\chi\chi}^2(\dot{\phi}^2+\dot{\phi}-2)+f_{\chi}\left(
8\dot{\phi}^2+3(\dot{\phi}-2)(2\dot{\phi}^2+\dot{\phi}-2)
f_{\chi\chi}\right)\right]Q_1^2\\
\nonumber
&+&12a^6\dot{\phi}^2f_{\chi\chi}\left[\left(3\ddot{\phi}^2
f_{\chi\chi}^2+\dot{H}\dot{\phi}^4(1-3f_{\chi\chi})-3
\dot{\phi}^2(2H\dot{\phi}+\ddot{\phi})f_{\chi}\right)Q_1+
9\dot{H}\dot{\phi}^4f_{\chi\chi} Q_3\right]Q_1\\
\nonumber
&-&18a^6\dot{\phi}^2f_{\chi}\left[f_{\chi}(4\dot{\phi}^2+
3(\dot{\phi}-1)^2(\dot{\phi}+2)f_{\chi\chi})+6f_{\chi\chi}
(\dot{\phi}-1)(H\dot{\phi}^2+\ddot{\phi}f_{\chi\chi})
\right] Q_1Q_3\\
\nonumber
&+&12a^6 \dot{\phi}^4f_{\chi\chi}\left[(\dot{\phi}^2-1)(
3Q_3Q_4+Q_5)-(\dot{\phi}^2+1)Q_1Q_4+3(\dot{H}(1-
\dot{\phi}^4)+H^2) Q_1Q_3\right]\\
\nonumber
&-&12 k^2 a^4 \dot{\phi}^4
(2Q_1+Q_3)Q_3+12k^2a^6\dot{\phi}^5f_{\chi}f_{\chi\chi}
Q_1Q_2-36k^2a^6 \dot{\phi}^4f_{\chi\chi}(2H+\dot{\phi}
f_{\chi})Q_2Q_3\\
\nonumber
&+&6a^3\dot{\phi}^4f_{\chi\chi}\left((2H-\dot{\phi}
f_{\chi}) P_3-6HP_1\right)Q_1-4k^2a^3\dot{\phi}^4
f_{\chi\chi}(P_3+k^2a^3Q_2)Q_2\\
\nonumber
&+&27a^6\dot{\phi}^2\left[6H\dot{\phi}^2f_{\chi} 
f_{\chi\chi}+(2\dot{\phi}^2+3(\dot{\phi}-1)^2f_{\chi\chi})
f_{\chi}^2+2\dot{H}\dot{\phi}^2f_{\chi\chi}(1+\dot{\phi}^2
(1-3f_{\chi\chi}))\right]Q_3^2\\
\nonumber
&+&12 a^3\dot{\phi}^4 f_{\chi\chi}(u_1\Upsilon_1+v_1
\Xi_1+w_1\Theta_1) \Bigg] \, .
\ea
This Hamiltonian is used for the constraint analysis after Eq. (\ref{T-Hamiltonian}). \\

{}

\end{document}